\documentclass[aps,twocolumn]{revtex4}

\usepackage{color}
\usepackage{graphicx}
\usepackage{amsmath}
\usepackage{amssymb}

\newcommand{\NOTE}[1]{{\color{black}{#1}}}

\newcommand{\MT}[4]{\left[\begin{array}{ccc}#1&\:&#2\\#3&\:&#4\end{array}\right]}

\begin{document}
\title{Low-Loss High-Fidelity Frequency-Mode Hadamard Gates Based on Electromagnetically Induced Transparency}

\author{Kao-Fang Chang$^1$}
\author{Ta-Pang Wang$^1$} 
\author{Chun-Yi Chen$^1$} 
\author{Yi-Hsin Chen$^{2,5}$}
\author{Yu-Sheng Wang$^1$}
\author{Yong-Fan Chen$^{3,5,\ast}$}
\author{Ying-Cheng Chen$^{4,5}$}
\author{Ite A. Yu$^{1,5,}$\footnotemark
\footnotetext{yu@phys.nthu.edu.tw; yfchen@mail.ncku.edu.tw}
}

\affiliation{
$^{1}$Department of Physics, National Tsing Hua University, Hsinchu 30013, Taiwan \\
$^{2}$Department of Physics, National Sun Yat-sen University, Kaohsiung 80424, Taiwan \\ 
$^{3}$Department of Physics, National Cheng Kung University, Tainan 70101, Taiwan \\
$^{4}$Institute of Atomic and Molecular Sciences, Academia Sinica, Taipei 10617, Taiwan \\
$^{5}$Center for Quantum Technology, Hsinchu 30013, Taiwan
}

\date{\today}

\begin{abstract}
A frequency beam splitter (FBS) with the split ratio of 0.5 or 1 can be used as the frequency-mode Hadamard gate (FHG) for frequency-encoded photonic qubits or as the quantum frequency converter (QFC) for frequency up or down conversion of photons. Previous works revealed that all kinds of the FHG or QFC operating at the single-photon level had overall efficiency or output-to-input ratio around 50\% or less. In this work, our FHG and QFC are made with the four-wave mixing process based on the dual-$\Lambda$ electromagnetically induced transparency scheme. We achieved an overall efficiency of 90$\pm$4\% in the FGH and that of 84\% in the QFC using coherent-state single photons, both of which are the best up-to-date records. To test the fidelity of the FBS, we propose a novel scheme of Hong-Ou-Mandel interference (HOMI) for quantum process tomography. The fidelity indicated by the HOMI's $g^{(2)}$ measurement of the FHG is 0.99$\pm$0.01. Such low-loss high-fidelity FHG and QFC or FBS with the tunable split ratio can lead to useful operations or devices in long-distance quantum communication.
\end{abstract}

\maketitle

%%%%%%%%%%%%%%%%%%%%%%%%%%%%%%%%%%%%%%%%%%%%%%%
%%%%%%%%%%%%%%%%%%%%%%%%%%%%%%%%%%%%%%%%%%%%%%%
\newcommand{\FigOne}{
	\begin{figure}[t]
	\includegraphics[width=8.75cm]{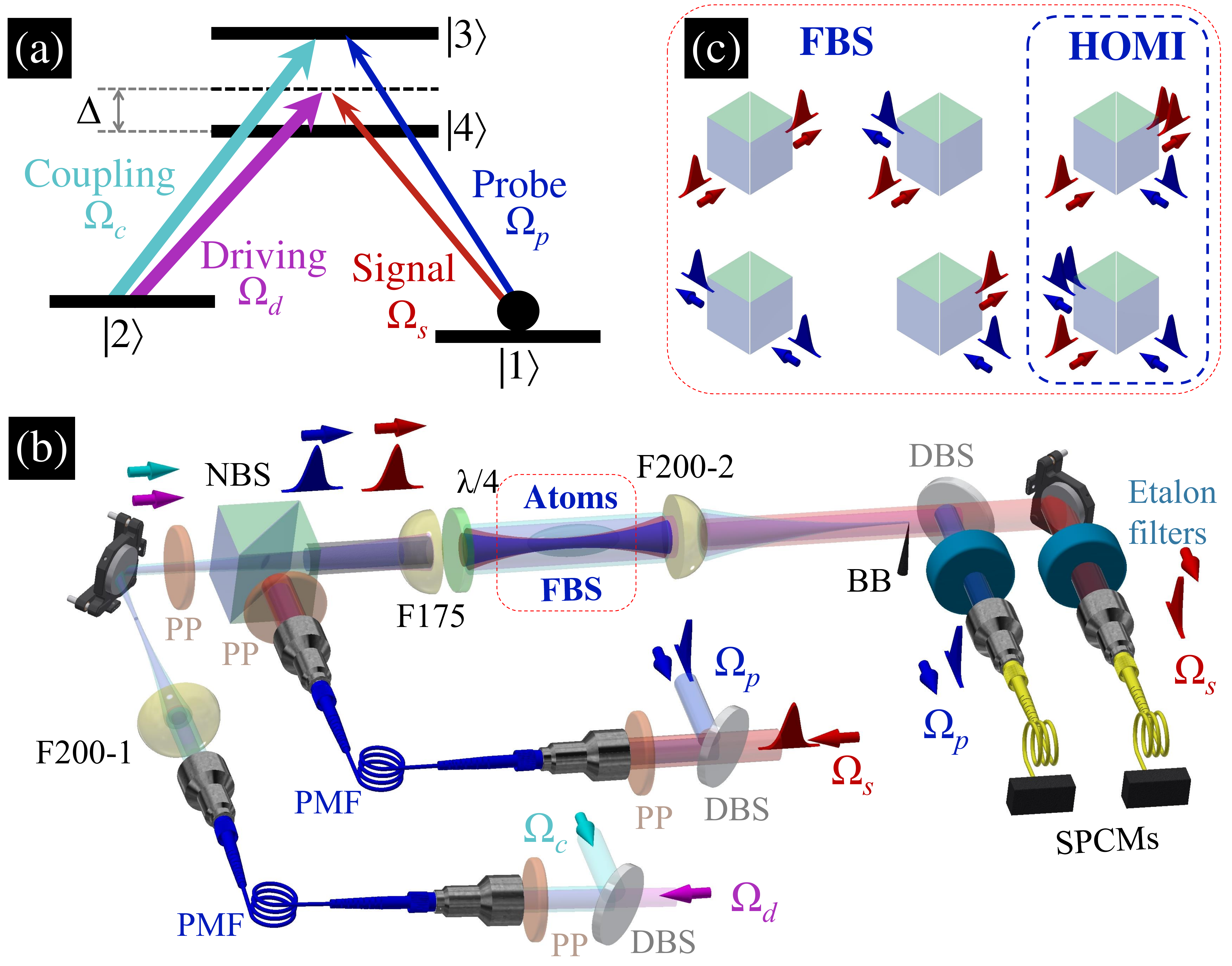}
	\caption{Transition diagram, experimental setup, and operation of FBS.
(a) Relevant energy levels and laser excitations in the experiment. $|1\rangle$, $|2\rangle$, $|3\rangle$, and $|4\rangle$ represent the states $|5S_{1/2},F=1,m=1\rangle$, $|5S_{1/2},F=2,m=1\rangle$, $|5P_{3/2},F=2,m=2\rangle$, and $|5P_{1/2},F=2,m=2\rangle$ of $^{87}$Rb atoms, respectively. The coupling ($\Omega_c$) and probe ($\Omega_p$) fields with the wavelength of 780 nm drove the transitions resonantly. The driving ($\Omega_d$) and signal ($\Omega_s$) fields with the wavelength of 795 nm drove the transitions with a detuning of $\Delta$. (b) Schema of the experimental setup. DBS: dichroic beam splitter; PP: polarizer or polarizing beam splitter with half-wave plate; PMF: polarization-maintained optical fiber; PBS: polarizing beam splitter; NBS: non-polarizing beam splitter with $T/R$ = 90/10; F200-1, F175, F200-2: lenses with focal lengths of 200, 175, and 200 mm; $\lambda$/4: quarter-wave plate; BB: beam block; SPCM: single-photon counting module. (c) Illustration of 50/50 FBS. The left four diagrams depict that a 780 nm (or 795 nm) photon arriving to the input can result in either a 780 nm photon or a 795 nm photon with the equal probability at the output. The right two diagrams depict that a 780 nm and a 795 nm photons simultaneously arriving to the input can produce two photons of the same wavelength at the output due to the Hong-Ou-Mandel interference.
}
	\label{fig:Transition}
	\label{fig:Setup}
	\label{fig:FBSop}
	\end{figure}
}
%%%%%%%%%%%%%%%%%%%%%%%%%%%%%%%%%%%%%%%%%%%%%%%
\newcommand{\FigTwo}{
	\begin{figure}[!]
	\includegraphics[width=7cm]{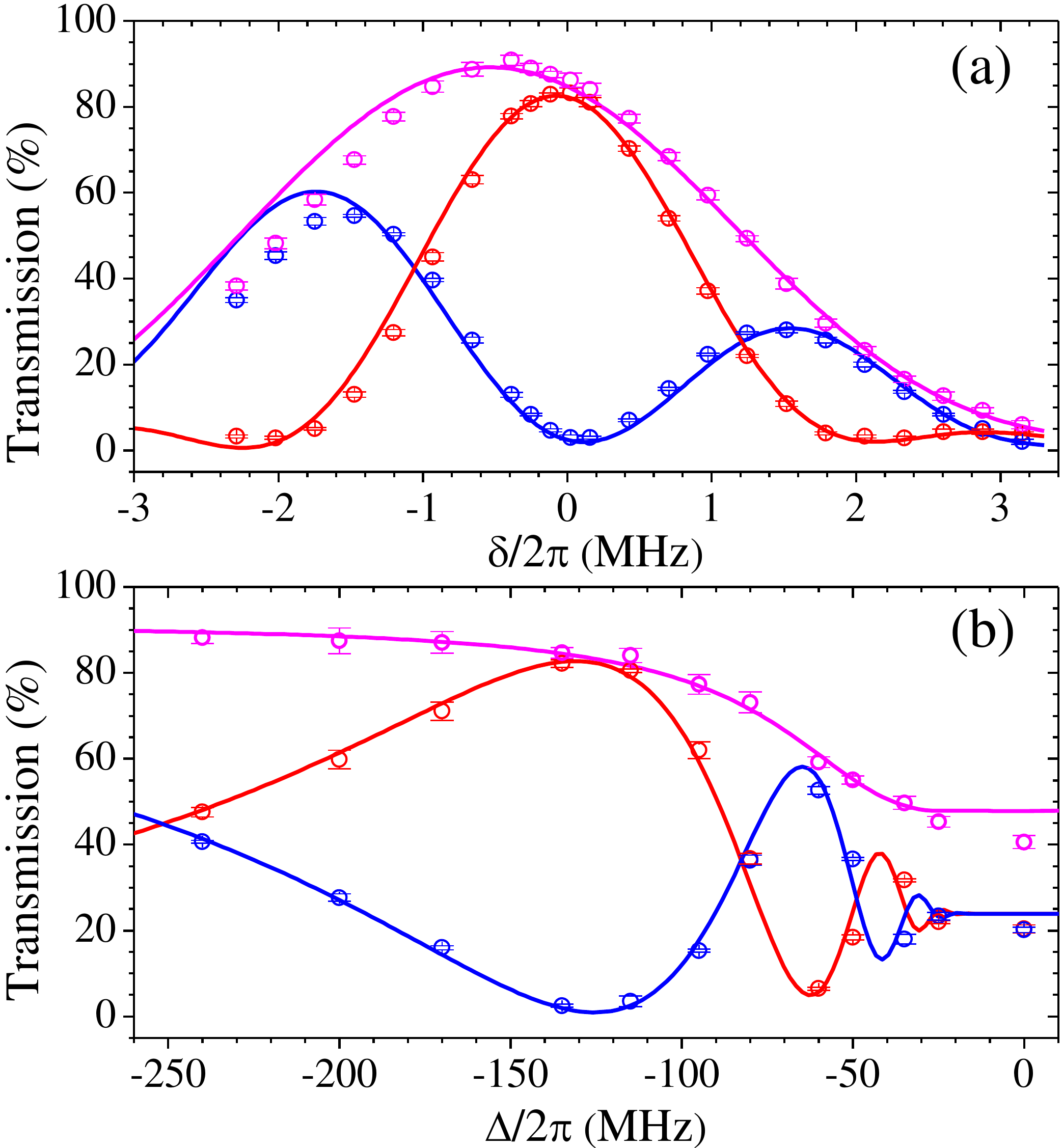}
	\caption{Tuning the split ratio of frequency beam splitter.
(a) At one-photon detuning $\Delta$ = $-2\pi\times 135$ MHz, transmissions as functions of two-photon detuning $\delta$, where $\delta = 0$ is defined by the maximum transmission of the output signal pulse, but not by making the two-photon Raman transition of $|1\rangle \rightarrow |2\rangle$ resonant. (b) At $\delta = 0$, transmissions as functions of $\Delta/(2\pi)$. In (a) and (b), only the 780 nm probe pulse with the $e^{-2}$ full width of 3.0 $\mu$s was present at the input. Blue, red, and magenta circles are the experimental data of 780 nm probe and 795 nm signal output transmissions, and their total transmission, respectively. Solid lines are the theoretical predictions calculated with $\Omega_c = \Omega_d$ = 3.0$\Gamma$, $\alpha$ (OD) = 130, and $\gamma = 3\times10^{-3}$$\Gamma$, which were experimentally determined by the method illustrated in \NOTE{Sec.~III of the Supplemental Material.} The asymmetry between positive and negative values of $\delta$ is due to the existence of a phase mismatch in the experimental system.
}
	\label{fig:SplitRatio}
	\end{figure}
}
%%%%%%%%%%%%%%%%%%%%%%%%%%%%%%%%%%%%%%%%%%%%%%%
\newcommand{\FigThree}{
	\begin{figure}[!]
	\includegraphics[width=8.5cm]{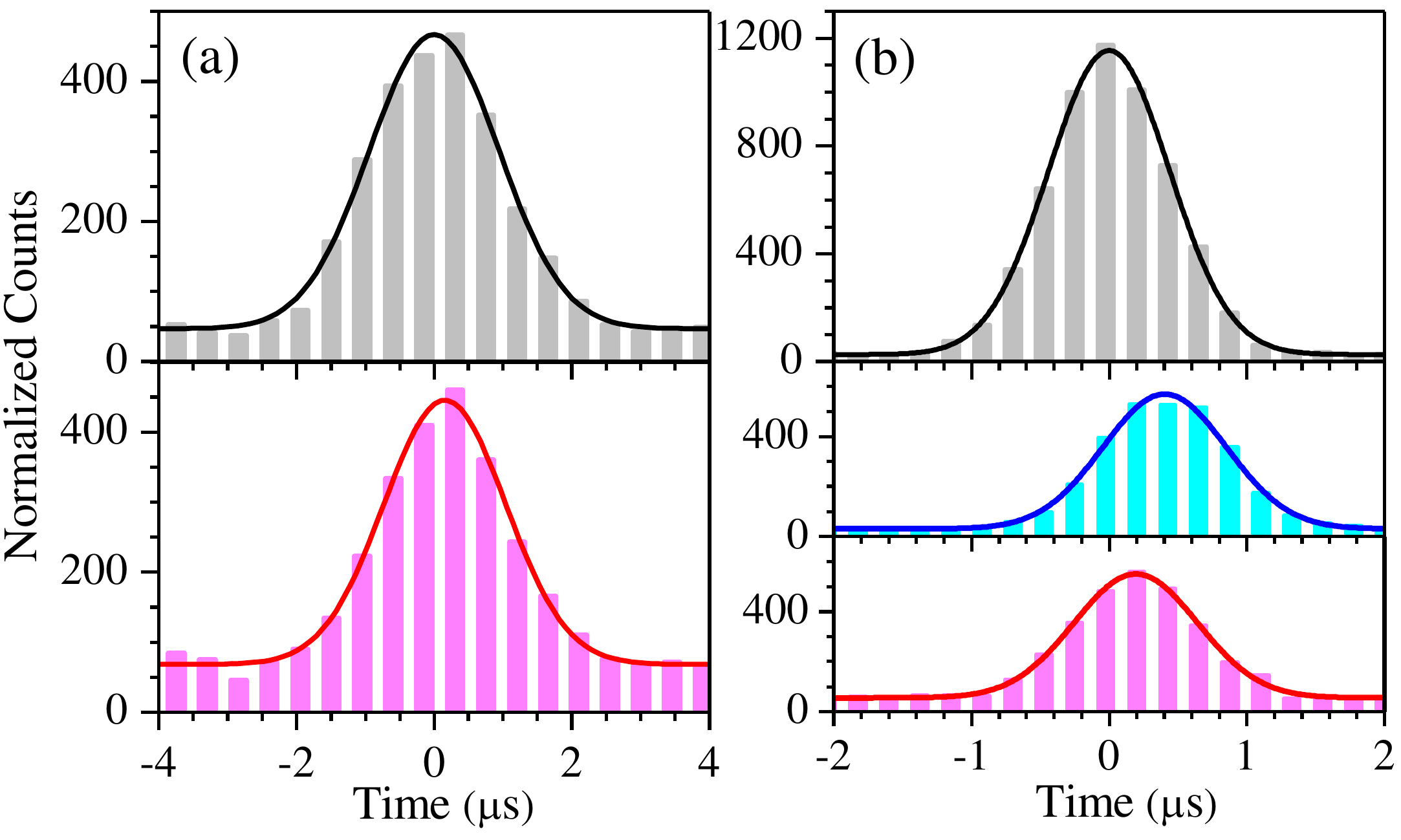}
	\caption{Single-photon operations of frequency converter in (a) and frequency beam splitter in (b).
In (a), the number of input photon per pulse was 0.68, and the split ratio is nearly 1 set by $\Delta/(2\pi)$ = $-125$ MHz. Top: counts of 780 nm input photons; bottom: those of 795 nm output photons. In (b), the number of input photon per pulse was 1.0, and the split ratio was about 0.5 set by $\Delta/(2\pi)$ = $-210$ MHz. Top: counts of 780 nm input photons; middle and bottom: those of 780 nm and 795 nm output photons. In (a) or (b), the width of time bin for SPCM counts was 450 ns or 225 ns, and the data were the results of 24,000 or 32,000 measurements. All of black, red, and blue lines are the Gaussian best fits. Excluding the baseline count, the ratios of output to input photon numbers are 84$\pm$4\% in (a) and 90$\pm$4\% in (b).
}
	\label{fig:SinglePhoton}
	\end{figure}
}
%%%%%%%%%%%%%%%%%%%%%%%%%%%%%%%%%%%%%%%%%%%%%%%
\newcommand{\FigFour}{
	\begin{figure}[!]
	\includegraphics[width=7cm]{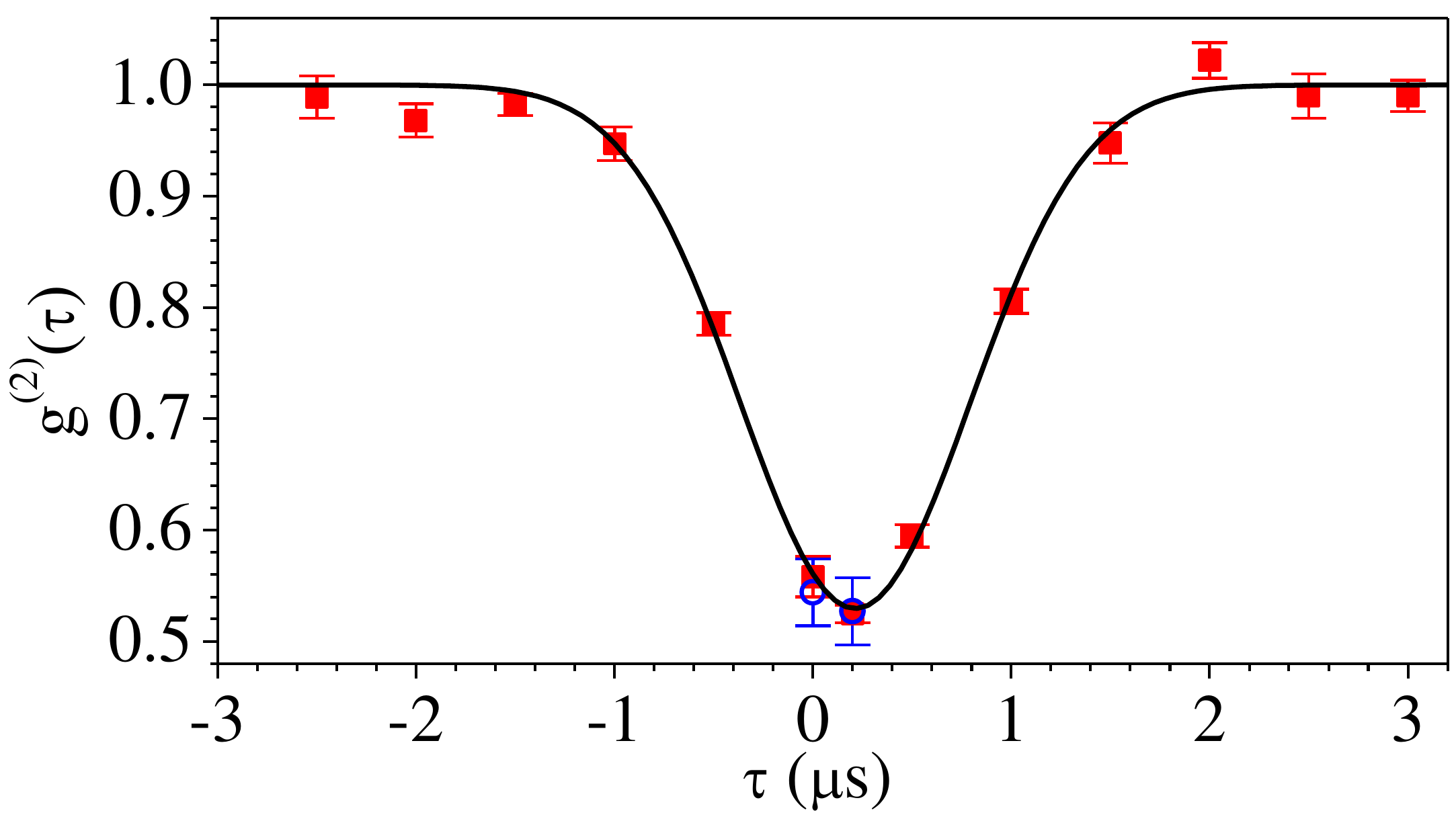}
	\caption{Measurement of cross correlation function $g^{(2)}$ in the Hong-Ou-Mandel interference.
$g^{(2)}$ is plotted against the delay time between two input pulses, $\tau$. The experimental condition was very similar to that in \NOTE{Fig.~S2 of the Supplemental Material.} Blue circles are the experimental data taken at $\Delta/(2\pi)$ = $-215$ MHz with two single-photon input pulses of 780 nm and 795 nm. Each circular data point, in which the contribution from background counts is removed, is the result of 28,800 measurements. Red squares are the experimental data taken at $\Delta/(2\pi)$ = $-205$ MHz with two 5-photon input pulses. Each square data point, corrected for the saturation effect of SCPM's gain and the background contribution, is the result of 9,600 measurements. The two values of $\Delta$, which set the split ratio of FBS to about 0.5, are different due to the day-to-day variation of OD. The $e^{-2}$ full width of input pulses was 1.7 $\mu$s. The time windows for SPCM counts of the blue circular and red square data points were 1.8 and 8.0 $\mu$s, respectively. Solid line is the best fit of a Gaussian function with the $e^{-2}$ full width of 2.3 $\mu$s and the minimum $g^{(2)}$ of 0.53.
}
	\label{fig:HOMI}
	\end{figure}
}
%%%%%%%%%%%%%%%%%%%%%%%%%%%%%%%%%%%%%%%%%%%%%%%
\newcommand{\FigFive}{
	\begin{figure}[t]
	\includegraphics[width=7cm]{_Fig5}
	\caption{{\bf A 50/50 frequency beam splitter operating with classical pulses.}
(a) At $\Delta/(2\pi)$ = $-205$ MHz, $\alpha$ (OD) = 110, and $\Omega_c$ = $\Omega_d$ = 3.0$\Gamma$, the FWM process converted the input 780 nm probe (black) pulse to the output 795 nm signal (red) and 780 nm probe (blue) pulses with the transmissions of 46\% and 45\%, respectively. The delay time between the red (or blue) and black lines is 0.21 (or 0.42) $\mu$s. (b) Under the same experimental condition, the FWM process converted the input 795 nm (black) pulse to the output 795 nm (red) and 780 nm (blue) pulses with the transmissions of 53\% and 38\%, respectively. The delay time between the red (or blue) and black lines is $-$0.04 (or 0.20) $\mu$s.
}
	\label{fig:FBS}
	\end{figure}
}
%%%%%%%%%%%%%%%%%%%%%%%%%%%%%%%%%%%%%%%%%%%%%%%
%%%%%%%%%%%%%%%%%%%%%%%%%%%%%%%%%%%%%%%%%%%%%%%

Quantum information or wave functions is commonly encoded in photons' polarization or spatial mode. As compared with these two kinds of photonic qubits, frequency-encoded qubits are not only more stable over long transmission distances but also more robust against birefringent materials \cite{OE2011, PRA2014, SSL2014, NPhot2016, Optica2017, PRL2018}. Among quantum logic operations, the Hadamard gate is an essential component. A beam splitter is exactly the Hadamard gate for spatial-mode qubits. In the context of frequency-encoded photonic qubits, a frequency beam splitter (FBS) is the Hadamard gate. In this work, we demonstrate a FBS with a tunable split ratio, where the split ratio is the ratio of photon number in one output frequency mode to total output photon number. At the split ratio of 0.5, a FBS, i.e., a 50/50 FBS, can be employed as the frequency-mode Hadamard gate (FHG). At the split ratio of 1, a FBS can be utilized as the quantum frequency converter (QFC), which coherently converts a photonic qubit from one frequency or wavelength to another. 

To date, all kinds of FHG and QFC operating at the single-photon level had output-to-input ratios or overall efficiencies (including decay due to propagation or insertion loss in media, input coupling efficiency, frequency conversion efficiency, etc.) around 50\% or less \cite{NPhot2016, PRL2018, FWM3, FWM4, FWM5, FWM9, FWM10, FWM11, FWM12, FWM13, FWM14, FWM15}. Most of these works suffered large insertion loss induced by media, which not only reduces the output-to-input ratio but also may lead to additional quantum noise. Here, our low-loss FBS is made with the four-wave mixing (FWM) process based on the dual-$\Lambda$ electromagnetically induced transparency (EIT) scheme \cite{ReviewEIT, EIT-FWM1, EIT-FWM2, EIT-FWM3, FWM1996, FWM2002, FWM2014, FWM2016}. Using the transition scheme depicted in Fig.~\ref{fig:Transition}(a), we converted a coherent-state single photon in the 780 nm mode to another photon in the superposition of 780 nm and 795 nm modes, and demonstrated that the FHG has an output-to-input ratio of 90$\pm$4\%. Furthermore, we performed the QFC from 780 to 795 nm with light pulses of photon number less than one, and achieved an output-to-input ratio of 84$\pm$4\%. Both output-to-input ratios are the best up-to-date records.

To test the fidelity of a quantum device or operation, one should perform quantum process tomography \cite{QPT1, QPT2, QPT3, QPT4}. We propose a novel method for quantum process tomography using Hong-Ou-Mandel interference (HOMI). \cite{HOMI1, HOMI2, HOMI3, HOMI4, HOMI5}. The HOMI is a two-photon phenomenon, in which one two-mode wave function formed by the two outputs of the FBS interferes with another. In the HOMI measurement of our FHG, the value of normalized cross correlation function, $g^{(2)}$, reveals that the fidelity is 0.99$\pm$0.01. To our knowledge, this is the first time that the HOMI is used for quantum process tomography of a quantum logic gate. The result of high fidelity also indicates that the single-photon quantum state is well preserved in the dual-$\Lambda$ EIT scheme. The EIT mechanism is universal and can work for various media \cite{EIT1, EIT2, EIT3, EIT4, EIT5}. Hence, the high-fidelity and low loss FBS reported here can be readily applied to systems of the optical depth and decoherence rate similar to those in this work.

\FigOne

Our experiment was carried out with laser-cooled $^{87}$Rb atoms \cite{CigarMOT, OurPRA2013, QM2013}. Figure~\ref{fig:Setup}(b) shows the schema of experimental setup. In the photon-atom coupling scheme as depicted in Fig.~\ref{fig:Transition}(a), the 780 nm probe and coupling fields form the first EIT configuration under the one-photon resonance; the 795 nm signal and driving fields form the second one with a large one-photon detuning, $\Delta$. The coupling and driving fields were strong quasi-cw light. The probe and signal fields were weak classical pulses, or coherent-state single- or few-photon pulses. Other details of the experimental system can be found in \NOTE{Sec.~I of the Supplemental Material.}

To characterize our experimental system and verify measurement outcomes, we made theoretical predictions with the optical Bloch equations (OBEs) of density-matrix operator and the Maxwell-Schr\"{o}dinger equations (MSEs) of light fields, which can be found in \NOTE{Sec.~II of the Supplemental Material.} In these equations and thorough the paper, $\Omega_c$, $\Omega_d$, $\Omega_p$, and $\Omega_s$ denote the Rabi frequencies of the coupling, driving, probe, and signal fields, $\delta$ is the two-photon detuning of the Raman transition between two ground states $|1\rangle$ and $|2\rangle$, $\gamma$ represents the ground-state decoherence rate, $\Gamma$ denotes the spontaneous decay rate of the excited states $|3\rangle$ and $|4\rangle$ which is about $2\pi\times$6 MHz in our case, and $\alpha$ is the optical depth (OD) of the medium. The measurements that determined $\Omega_c$, $\Omega_d$, $\gamma$, and $\alpha$ in the experiment are illustrated in \NOTE{Sec.~III of the Supplemental Material.}

The split ratio here is defined as the ratio of 795 nm output photon number to total output photon number under the condition that only the 780 nm photons are present at the input. Tuning the split ratio of FBS can be done by varying either two-photon detuning $\delta$ or one-photon detuning $\Delta$. In this study of split ratio, only the 780 nm probe pulse of classical light was present at the input, and $\Omega_c = \Omega_d$. A part of the 780 nm input pulse was converted to the 795 nm signal pulse at the output, and the remaining became the 780 nm output pulse. Figures~\ref{fig:SplitRatio}(a) and \ref{fig:SplitRatio}(b) show the energy transmissions of 780 nm and 795 nm output pulses as functions of the two-photon detuning $\delta$ and the one-photon detuning $\Delta$, respectively. One can see that using $\delta$ to tune the split ratio can suffer a larger loss, and using $\Delta$ is more efficient. In Fig.~\ref{fig:SplitRatio}(b), the split ratio can be tuned from 1 to 0.5 or smaller with $|\Delta|/(2\pi) \geq$ 130 MHz. The total energy transmission of 780 nm and 795 nm output pulses is 85\% (or 88\%) at the split ratio equal to 0.97 (or 0.54). A smaller split ratio results in a higher total transmission. In both figures, the apparent phenomenon of oscillation indicates that the underlying mechanism of FWM involves with the interference effect \cite{FWM2002, FWM2014}. The theoretical predictions were calculated by numerically solving OBEs and MSEs with the experimentally-determined parameters of $\alpha$, $\Omega_c$, $\Omega_d$, and $\gamma$ \cite{OurPRA2013, QM2013}. Consistency between the experimental data and theoretical predictions is satisfactory.

\FigTwo

To test whether the scheme of our FBS can also work well at the single-photon level, we performed the measurements with coherent-state pulses of photon number equal to or less than 1. Two etalon filters were installed to provide the extinction ratio of 43 dB. The etalons, together with the scheme of spatial filter (see the third paragraph in \NOTE{Sec.~I of the Supplemental Material),} can effectively block the strong coupling and driving light from entering single-photon counting modules (SPCMs). Two Excelitas SPCM-AQRH-13-FC were used to detect the 780 nm and 795 nm output photons. The collection efficiencies (including the SPCM's quantum efficiency) of the 780 nm and 795 nm photons were about 0.13 and 0.12 for the data in Fig.~\ref{fig:SinglePhoton}(a) [0.17 and 0.12 for those in Fig.~\ref{fig:SinglePhoton}(b)]. We had another SPCM at the input to monitor the input photon number. All of the photo multiplier tubes used in the measurements of classical-light pulses and the SPCMs used in those of single-photon or few-photon pulses were calibrated to account for different detection efficiencies between the wavelengths or between the detectors. In each SPCM's counting, it took 0.15 s to replenish cold atoms, switch off the MOT, perform the temporal dark-MOT, and optically pump all population to a single Zeeman state, before the input pulse was fired.

At the split ratio of 1, the FBS acts like a coherent wavelength converter transforming 780 nm photons completely into 795 nm photons. Figure~\ref{fig:SinglePhoton}(a) shows SCPM counts of input and output photons as functions of time. The best fit of the data in Fig.~\ref{fig:SinglePhoton}(a) is consistent with the result of classical light shown by Fig.~S1(c) in \NOTE{Sec.~III of the Supplemental Material.} The baseline count was mainly contributed from the leakage of strong coupling or driving fields. Using the area below the best fit but excluding the baseline count, we determined the overall conversion efficiency from the 780 nm single photons to the 795 nm single photons or the output-to-input ratio is 84$\pm$4\%. 

The split ratio of 0.5 can make the 50/50 FBS or Hadamard gate for frequency-encoded photonic quits. In Fig.~\ref{fig:SinglePhoton}(b), SCPM counts of 780 nm input photons and those of 780 nm and 795 nm output photons are plotted against time. The best fits of the data in Fig.~\ref{fig:SinglePhoton}(b) are consistent with the results of classical light shown by \NOTE{Fig.~S2(a)} in \NOTE{Sec.~IV of the Supplemental Material.} The comparison between Figs.~\ref{fig:SinglePhoton}(a) and \ref{fig:SinglePhoton}(b) indicates that employing a narrower single-photon pulse can increase the amplitude-to-baseline ratio, while the output-to-input ratio was nearly intact. In our 50/50 FBS, the total transmission or ratio of total output photons to input photons is 90$\pm$4\%.

\FigThree

We have now made the FBS which can operate with single photons. In analogy to an ordinary BS, 780 nm (or 795 nm) input photons are reflected into 795 nm (or 780 nm) output photons and transmitted into 780 nm (795 nm) output photons by our FWM-based FBS, with the split ratio defined by the ratio of reflected output photon number to total output photon number. Figure~\ref{fig:FBSop}(c) illustrates the operation of 50/50 FBS. The next question is whether this FBS can be suitable for quantum information processing. To answer the question, fidelity $F$ is the important issue and can be determined by the following formula \cite{Fidelity1, Fidelity2, PRL2018}:
\begin{equation}
	F = \frac{\left| {\rm Tr} \left[ \hat{V}^\dagger \hat{U} \right] \right|^2}{4T},
\label{eq:fidelity}
\end{equation}
where $\hat{U}$ represents the operator of an ideal BS, $\hat{V}$ represents the operator of FWM-based FBS in the case here, ${\rm Tr}[...]$ means the operation of trace, and $T$ is the total transmission or success probability of $\hat{V}$. 

Considering the FBS, we define $t_1$ ($t_2$) and $r_1$ ($r_2$) as the transmission and reflection coefficients of input 1 (input 2), and $\phi_1$ ($\phi_2$) as the phase difference between the reflected and transmitted outputs. The general expression of $\hat{V}$ is given by
\begin{equation}
	\MT{t_1}{r_2 e^{i\phi_2}}{r_1 e^{i\phi_1}}{t_2}.
\label{eq:Vmatrix}
\end{equation}
The split ratio of two inputs are determined by $r_1^2/(t_1^2+r_1^2)$ and $r_2^2/(t_2^2+r_2^2)$. In reality, the ground-state decoherence rate $\gamma$ in the experimental system was not negligible, making two inputs produce different split ratios and $\phi_1 \neq \phi_2$. An ideal 50/50 BS must have $\phi_1 + \phi_2 = \pi$. Corresponding to $\hat{V}$ in Eq.~(\ref{eq:Vmatrix}) of a realistic BS with two split ratios close to 0.5, $\hat{U}$ of the ideal 50/50 BS can be written as \cite{BS1, BS2}
\begin{equation}
	\hat{U} = \frac{1}{\sqrt{2}} \MT{1}{e^{i(\pi-\Delta\phi)/2}}{e^{i(\pi+\Delta\phi)/2}}{1},
\end{equation}
where $\Delta\phi \equiv \phi_1 - \phi_2$. The derivation in \NOTE{Sec.~V of the Supplemental Material} shows that the fidelity of $\hat{V}$ is
\begin{equation}
	F = \frac{1}{2}+ \frac{\bar{t}\bar{r}}{T} \sin\left( \frac{\phi}{2} \right),
\label{eq:FBSfidelity}
\end{equation}
where $\bar{t} = (t_1 + t_2)/2$, $\bar{r} = (r_1 + r_2)/2$, $T = \bar{t}^2 + \bar{r}^2$, and $\phi = \phi_1+\phi_2$. 

\FigFour

According to Eq.~(\ref{eq:FBSfidelity}), one can immediately see that the phase $\phi$ approaching to $\pi$ can make a high-fidelity FBS. To determine $\phi$, we employed the Hong-Ou-Mandel interference (HOMI) \cite{HOMI1, HOMI2, HOMI3, HOMI4, HOMI5}, and measured the normalized cross correlation function, $g^{(2)}$, between two outputs of the FBS. A simple example, to explain why the HOMI measurement can determine $\phi$ of Eq.~(\ref{eq:FBSfidelity}), is illustrated in \NOTE{Sec.~VI of the Supplemental Material.} In the HOMI, it is well known that, with a 50/50 BS in the ideal condition, two Fock-state single photons results in $g^{(2)} = 0$ \cite{HOMI1, HOMI2}, and two phase-uncorrelated coherent-state single photons results in $g^{(2)}$ = 0.5 \cite{HOMI3, HOMI4, HOMI5}. Here, we sent two pulses to the two input ports of 50/50 FBS in the HOMI measurement. Each pulse consisted of a coherent-state single photon or few photons. The wavelength of one pulse was 780 nm and that of the other was 795 nm. Since the two pulses had the same mean photon number and were phase-uncorrelated, the derivation in \NOTE{Sec.~VII of the Supplemental Material} shows that $g^{(2)}$ of the two output ports is given by
\begin{equation}
	g^{(2)} = 1 + \frac{2 t_1 t_2 r_1 r_2}{(t_1^2 + r_2^2)(t_2^2 +r_1^2)} \cos\phi.
\label{eq:g2}
\end{equation}
In Fig.~\ref{fig:HOMI}, $g^{(2)}$ is plotted against the delay time between the two input pulses. The minimum $g^{(2)}$ is $0.53\pm0.03$ and occurs at the delay time of 200 ns. Based on the data shown in \NOTE{Figs.~S2(a) and S2(b) of  the Supplemental Material,} this 200-ns delay time is expected. The two-photon event of both photons from two input ports transmitting through the FBS, and that of both photons being reflected by the FBS were nearly indistinguishable under such delay time. Right before taking the data in Fig.~\ref{fig:HOMI}, we measured the data similar to those in \NOTE{Figs.~S2(a) and S2(b),} and found $t_1^2 = 46\%$, $r_1^2 = 46\%$, $t_2^2 = 51\%$, and $r_2^2 = 39\%$. The minimum $g^{(2)}$ and Eq.~(\ref{eq:g2}) result in $\cos\phi = -0.94(6)$. Finally, we use Eq.~(\ref{eq:FBSfidelity}) and the above values of $t_1$, $r_1$, $t_2$, $r_2$, and $\phi$ to determine $F$ = 0.99$\pm$0.01, indicating that the FWM-based FBS possesses excellent fidelity.

In conclusion, utilizing the EIT-based FWM process we experimentally demonstrated the FBS with coherent-state single-photon pulses. At the split ratio of nearly 1, the FBS converted all of 780 nm input photons to 795 nm output photons with the output-to-input ratio of 84$\pm$4\%, and can be employed as a quantum frequency converter. At the split ratio of 0.5, the FBS offered the output-to-input ratio of 90$\pm$4\%, and can be employed as the Hadamard gate for frequency-encoded photonic qubits. Both of the output-to-input ratios or overall efficiencies are the best up-to-date records. To test the fidelity of the FBS, we proposed a method of the HOMI-type quantum process tomography, in which one two-mode wave function formed by two outputs of the Hadamard gate interferes with another. The value of $g^{(2)}$ in the HOMI measurement indicates that the fidelity of our frequency-mode Hadamard gate is 0.99$\pm$0.01. To our knowledge, it is the first time to utilize the HOMI in the quantum process tomography. This low-loss high-fidelity FBS with the tunable split ratio can lead to useful devices or operations, such as entanglement swapping, multiplexing, etc., in long-distance quantum communication.

\section*{ACKNOWLEDGMENTS}

This work was supported by the Ministry of Science and Technology of Taiwan under Grant Nos. 106-2119-M-007-003 and 107-2745-M-007-001.

%%%%%%%%%%%%%%%%%%%%%%%%%%%%%%%%%%%%%%%%%%%%%%%
%%%%%%%%%%%%%%%%%%%%%%%%%%%%%%%%%%%%%%%%%%%%%%%

%%%%%%%%%%%%%%%%%%%%%%%%%%%%%%%%%%%%%%%%%%%%%%%
%%%%%%%%%%%%%%%%%%%%%%%%%%%%%%%%%%%%%%%%%%%%%%%
\end{document}